\documentclass[twocolumn,showpacs,pra]{revtex4}

\usepackage{graphicx}%
\usepackage{dcolumn}
\usepackage{amsmath,amssymb}
\usepackage{bm}

\newcommand{\bfr}{{\bf r}}

\begin{document}


\title[Short Title]{Stability of multiquantum vortices in dilute Bose-Einstein condensates}

\author{T.~P. Simula}
\author{S.~M.~M. Virtanen}%
\author{M.~M. Salomaa}%
\affiliation{Materials Physics Laboratory, 
Helsinki University of Technology\\
P.~O.~Box 2200 (Technical Physics), FIN-02015 HUT, Finland}

\date{\today}

\begin{abstract}
Multiply quantized vortices in trapped Bose-Einstein condensates are studied using the Bogoliubov theory. Suitable combinations of a localized pinning potential and external rotation of the system are found to energetically stabilize, both locally and globally, vortices with multiple circulation quanta. We present a phase diagram for stable multiply quantized vortices in terms of the angular rotation frequency and the width of the pinning potential. We argue that multiquantum vortices could be experimentally created using these two expedients.
\end{abstract}

\pacs{03.75.Fi, 05.30.Jp, 67.40.Db}
\maketitle

\section{Introduction}
Bose-Einstein condensation in trapped atomic gases was realized and observed in 1995 \cite{Anderson1995a,Bradley1995a,Davis1995b}. Later, those pioneering experiments were followed by the creation of singly quantized vortices \cite{Matthews1999b} and vortex lattices \cite{Madison2000a,Abo2001a} in such systems. Particularly after the creation of vortices in dilute atomic Bose-Einstein condensates (BECs), there has been pronounced interest in studying vortex configurations \cite{FetterReview} due to their inherent connection with superfluidity. Especially, the stability of vortices in BECs has been subject to intensive research \cite{Rokhsar1997a,Feder1999a,Castin1999b,Garcia-Ripoll1999a,Pu1999a,Tomppa1999a,Isoshima1999b,Garcia-Ripoll2000b,Svidzinsky2000a}.

Singly quantized vortices in nonrotating, harmonically trapped BECs are locally energetically unstable within the Bogoliubov approximation in the sense that their excitation spectra contain anomalous negative-energy eigenmodes having a positive norm \cite{Dodd1997a,FetterReview}. If dissipation in the system is not negligible, this local instability implies vortices to spiral out of the condensate \cite{Rokhsar1997a}. However, the anomalous modes are shifted to positive energies (\emph{w.r.t.} the condensate energy) under sufficient rotation of the system \cite{FetterReview}. Furthermore, self-consistent finite-temperature theories predict singly quantized vortices to be locally stable even in the absence of external rotation \cite{Tomppa1999a,Virtanen2001a,Virtanen2001b}. In fact, a sufficient criterion for the existence of single-quantum vortices in dilute BECs seems to be their global stability, \emph{i.e.}, their having a lower free energy compared to the nonvortex state. This can also be quaranteed by adequate rotation of the system.
 
Presently, many of the intriguing properties previously explored in conventional superfluid systems are being witnessed in dilute BECs. In addition to singly quantized vortices and vortex lattices, multiquantum vortices have been observed in thin films of superfluid $^4$He \cite{Marston} and in bulk rotating $^3$He-A \cite{Blaauwgeers2000a}. In superconductors, they can be stabilized with the use of holes, antidots, or columnar defects as the pinning centers \cite{Baert,Moshalkov,Braverman}. Contrary to the situation in Helium superfluids and superconductors, multiquantum vortices remain unobserved in dilute BECs. This is presumably due to their generic instability against dissociation into an array of single quantum vortices \cite{Butts1999a,Garcia-Ripoll1999a,Castin1999b}. Consequently, an external rotation of the system would rather increase the number of vortices than nucleate a multiply quantized vortex. Nevertheless, in suitable anharmonic traps multiquantum vortices have been argued to be (globally) energetically favorable with respect to vortex lattices \cite{Lundhpp}. 

In this paper, we demonstrate that by using vortex pinning \cite{Matthews1999b,Tomppa1999a}, one could stabilize vortices with multiple circulation quanta in harmonically trapped dilute BECs subject to an external rotation. We show that the multiply quantized vortex states become locally stable and energetically advantageous when suitable external rotation and localized pinning potentials are employed. The required pinning of vortices could be realized by using an intense, blue detuned laser beam \cite{Raman1999a}. We argue that using this scheme, multiquantum vortices could be experimentally created.   

\section{Mean-field approximation}
The low-temperature dynamics of the condensate is described by the time-dependent Gross-Pitaevskii equation \cite{Dalfovo1999a}
\begin{equation}
i\hbar\partial_t\Phi(\bfr,t)=\mathcal{H}(\bfr)\Phi(\bfr,t)+ g|\Phi(\bfr,t)|^2\Phi(\bfr,t).
\label{GPtime}
\end{equation}
The condensate wavefunction $\Phi(\bfr,t)$ is normalized according to $\int|\Phi(\bfr)|^2 {\rm d}\bfr = N$, where $N$ is the total number of particles. Above, the strength of interactions $g=4\pi\hbar^2 a/M$ as expressed in terms of the mass $M$ of the atoms and the $s$-wave scattering length $a$. The effective single-particle Hamiltonian
\begin{equation}
\mathcal{H}(\bfr)=-\frac{\hbar^2}{2M}\nabla^2+V_{\rm tr}(\bfr) + V_{\rm pin}(\bfr) +\bm{\Omega}\cdot(\bfr\times i\hbar\nabla)
\label{Hami}
\end{equation}
contains, in addition to the kinetic energy term, external trap and pinning potentials $V_{\rm tr}(\bfr)$ and $V_{\rm pin}(\bfr)$, respectively, and the angular momentum term, arising from the rotation of the system at the angular velocity ${\bm\Omega}$.

Stationary solutions of the form $\Phi(\bfr,t)=\phi(\bfr)e^{-i\mu t/\hbar}$ to Eq.~(\ref{GPtime}) obey the time-independent Gross-Pitaevskii equation
\begin{equation}
\mu\phi(\bfr)=\mathcal{H}(\bfr)\phi(\bfr)+g|\phi(\bfr)|^2\phi(\bfr),
\label{GP}
\end{equation}
where $\mu$ denotes the chemical potential determined by the normalization condition for the condensate wavefunction. 

The solutions of the stationary GP equation are minimizers of the energy functional 
\begin{equation}
E=\int [\Phi^*(\bfr) \mathcal{H}(\bfr)\Phi(\bfr) +\frac{g}{2}|\Phi(\bfr)|^4] \rm{d}\bfr.
\label{Functional}
\end{equation}
By using Eq.~(\ref{GP}), the energy of the condensate may be written in the form
\begin{equation}
E=\mu N-\frac{g}{2}\int|\phi(\bfr)|^4 \rm{d}\bfr
\label{Free}
\end{equation}
in terms of the chemical potential and the condensate wavefunction.

Adding a small perturbation
\begin{equation}
\delta\Phi(\bfr,t)= \epsilon[u_q(\bfr)e^{-iE_qt/\hbar}+ v^*_q(\bfr)e^{iE_qt/\hbar}]e^{-i\mu t/\hbar}
\label{Deltaphi}
\end{equation}
to a stationary solution of Eq.~(\ref{GP}), reinserting it to the GP equation and neglecting terms of higher than linear order in $\epsilon$ yields equations equivalent to the usual Bogoliubov equations \cite{Fetter72}
\begin{subequations}
\begin{eqnarray}
\mathcal{L} u_q(\bfr) +g\phi^2(\bfr)v_q(\bfr)&=&E_qu_q(\bfr), \\ \label{Bogo1}
\mathcal{L} v_q(\bfr) +g\phi^{*2}(\bfr)u_q(\bfr)&=&-E_qv_q(\bfr)   \label{Bogo2}
\end{eqnarray}
\label{Bogo}
\end{subequations}
for quasiparticle amplitudes $u_q(\bfr),v_q(\bfr)$, and eigenenergies $E_q$; here $q$ labels the quasiparticle states. Above, $\mathcal{L}\equiv\mathcal{H}(\bfr)-\mu +2g|\phi(\bfr)|^2$, and the quasiparticle amplitudes must obey the normalization $\int[|u_q(\bfr)|^2-|v_{q'}(\bfr)|^2] {\rm d}\bfr=\delta_{qq'}$, manifesting the bosonic character of the excitations. The condition of local energetic stability of the solutions to the GP equation is that there exists no positive-norm quasiparticle excitations with negative energies $E_q$ in the spectrum of Eqs.~(\ref{Bogo}). 

In what follows, we consider a Bose-Einstein condensate radially confined by a harmonic trapping potential $V_{\rm tr}(\bfr)=\frac{1}{2}M\omega^2r^2$ in cylindrical coordinates $\bfr=(r,\theta,z)$. Here $\omega$ is the harmonic frequency of the trapping potential. The system is rotated in the plane perpendicular to the symmetry axis of the trap. We study the stability of rectilinear multiquantum vortex lines of the form $\phi(\bfr)=\phi(r)e^{im\theta}$, located along the rotation axis. The vorticity $m$ determines the number of circulation quanta in the multiquantum vortices. We consider pinning potentials of the form $V_{\rm pin}(\bfr)=A\Theta(R_{\rm pin}-r)$, where $\Theta$ denotes the unit step function and the amplitude $A \gg \mu$ \footnote{For simplicity, we use a step potential to approximate a Gaussian laser beam having a sharp boundary. The amplitude dependence of the potential is removed by choosing the amplitude $A$ such that effectively all particles are removed from the volume occupied by the pinning potential. In this limit of strong pinning, the difference between the steplike and Gaussian-shaped beams becomes negligible.}. For numerically solving Eqs.~(\ref{GP}) and (\ref{Bogo}), we use computational methods similar to those described in Refs.~\cite{Virtanen2001a,Virtanen2001b,Simula2001a}.

\section{Energetics of vortices}
In order to approach thermal equilibrium, a physical system gravitates to the state that minimizes its free energy. Consequently, multiquantum vortices may be observed in experiments, provided that their free energies can be made lower than those of any other configurations---especially those of vortex arrays. However, in the absence of pinning, the energy of a multiquantum vortex is greater than that of a collection of singly quantized vortices with the same vorticity $m$. As a consequence, multiquantum vortices have not yet been seen in dilute, harmonically trapped BECs.  

In the case of a Bose-Einstein condensate without a rotating drive, the free energy minimizing state is vortex-free. However, when the condensate is rotated, its response is to acquire angular momentum by nucleating vortices. Pre-eminently, there exists a critical rotation frequency $\Omega_{\rm c}$ at which the energy of the condensate containing a singly quantized vortex becomes equal to that of a nonvortex state. At higher frequencies, vortices begin to nucleate from the edge of the condensate and to form arrays in the condensate \cite{Butts1999a,Abo2001a}. 

In the absence of the pinning potentials in the condensate volume, there exists no global minima in the effective potential felt by a vortex for rotation frequencies $\Omega < \Omega_{\rm c}$ \cite{Jackson1999b}. By the effective potential we mean the energy of the system as a function of the position of the vortex in the condensate. Adding a localized pinning potential on the trap axis lowers locally the effective potential and can create a global minimum in the vicinity of the trap symmetry axis, thus enabling vortex occupation there. However, for $\Omega < \Omega_{\rm c}$, the condensate volume outside the pinning potential would still remain energetically disadvantageous for vortices. The presence of circulation quanta in the system does not change the situation due to the repulsive interaction between vortices \cite{Castin1999b}.   

In conclusion, for rotation frequencies $\Omega < \Omega_{\rm c}$, states containing vortices in the region between the edge of the pinning potential and the condensate boundary are not globally energetically stable. In this paper we restrict the study to rotation frequencies $\Omega < \Omega_{\rm c}$, and thus do not consider the energetics of nonaxisymmetric vortex states. Extending the analysis consistently to frequencies $\Omega > \Omega_{\rm c}$ would be an elaborate task due to the complications arising from different vortex array configurations, mutual interaction between vortices and spatial dependence of the vortex self-energy.

\section{Stability of multiquantum vortices}
\subsection{Global stability}
Mere rotation of the condensate is not sufficient for stabilizing multiquantum vortices in harmonically trapped Bose-Einstein condensates. In order to render multiquantum vortex structures globally stable, an additional pinning potential is required besides the rotating drive. 

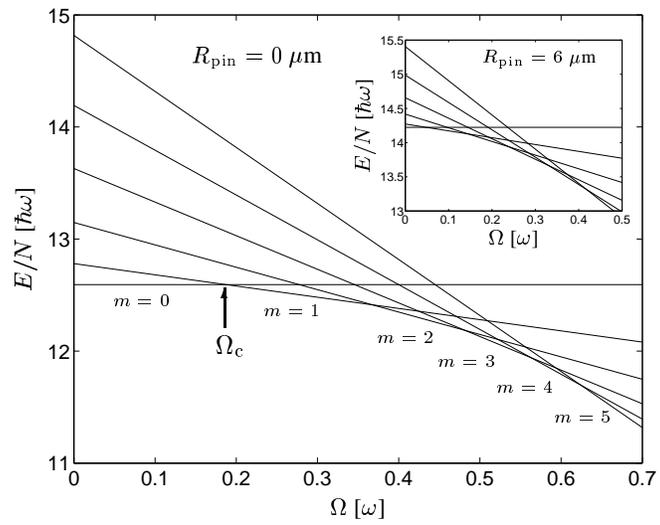
\begin{figure}[!h]
\caption{Free energies per particle as functions of the angular frequency $\Omega$ for circulation quanta $m=0,1,2,3,4$ and $5$. The radii of the pinning potentials used are 0 $\mu$m and 6 $\mu$m (inset). In the absense of the pinning potential, the energy of a singly quantized vortex becomes equal to that of a nonvortex state at the critical rotation frequency $\Omega_{\rm c}/\omega=0.19$. Notice the shifting of the intersections of the lines to lower values of $\Omega$ as $R_{\rm pin}$ is increased. In the results presented in this paper, $Na/a_{\rm ho}\approx 90$, where $a_{\rm ho}=(\hbar/M\omega)^{1/2}$ is the harmonic oscillator length, $\omega/2\pi=7.8$ Hz, and $N$ is the number of particles per harmonic oscillator length along the cylinder axis.} 
\label{Fig1}
\end{figure}

The computed energies of axisymmetric vortex states containing $m$ circulation quanta are shown in Fig.~\ref{Fig1} for $R_{\rm pin}=0$ $\mu$m and $R_{\rm pin}=6$ $\mu$m, as functions of the angular rotation frequency $\Omega$. In the absense of the pinning potential, the energy of a singly quantized vortex becomes equal to that of a nonvortex state at the critical rotation frequency $\Omega_{\rm c}$. Rotation does not change the energy of the nonvortex state $m=0$ because it has zero angular momentum. In the absense of the pinning potential, the multiquantum states are not the true energy minima for $\Omega>\Omega_{\rm c}$, due to the dissociation instability. However, when a pinning potential is added, the stabilizing frequencies of multiquantum vortices are shifted below $\Omega_{\rm c}$ (see the inset in Fig.~\ref{Fig1}), where the nonaxisymmetric states are not energetically favored. 

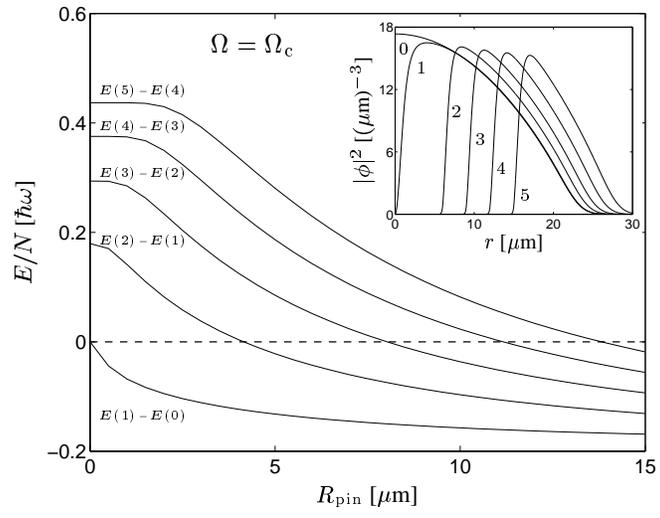
\begin{figure}[!h]
\caption{Energy differences $\Delta E(m)=E(m)-E(m-1)$  between consecutive multiquantum vortex states at  $\Omega=\Omega_{\rm c}$ as functions of the radius of the pinning potential $R_{\rm pin}$ for $m=1,2,3,4$ and $5$. The points $\Delta E(m)=0$ determine the minimum radii of the pinning potential for which multiquantum vortices with $m$ circulation quanta become energetically favorable. The subfigure shows stabilized, radial multiquantum density profiles for $m=0,1,2,3,4$ and $5$. The respective radii of the pinning potential $R_{\rm pin}(m)$ are chosen to be (0, 0, 6, 9, 12 and 15) $\mu$m.}
\label{Fig2}
\end{figure}

Figure \ref{Fig2} presents the energy differences $\Delta E(m)=E(m)-E(m-1)$ between adjacent multiquantum states at $\Omega=\Omega_{\rm c}$ as functions of the radius of the pinning potential. The intersections $\Delta E(m)=0$ in Fig.~\ref{Fig2} define the radii $R_{\rm pin}$ for which multiquantum configurations with $m$ circulation quanta become globally energetically stable at $\Omega=\Omega_{\rm c}$. In the inset, there are shown condensate density profiles for vortices with $m=0,\ldots,5$, stabilized by sufficiently wide pinning potentials. The rather large stabilizing values of $R_{\rm pin}$ compared to the core size of an unpinned vortex are partly due to the low rotation frequency applied, but also reflect condensate's rather strong tendency to distribute the vorticity uniformly throughout the system. 

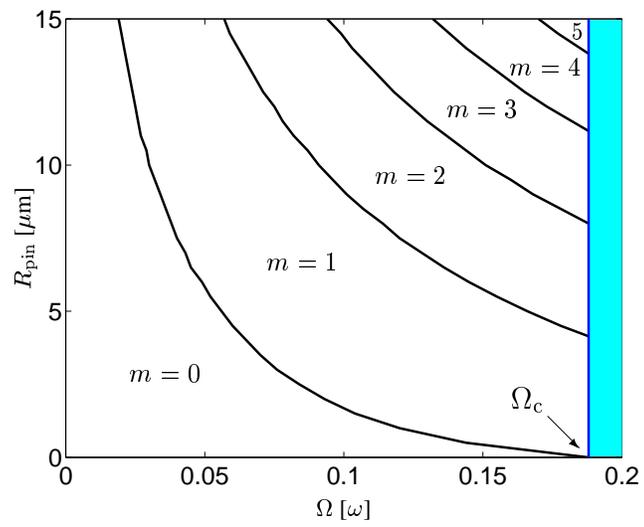
\begin{figure}[!h]
\caption{Phase diagram for the stability of multiquantum vortices in terms of the angular rotation frequency $\Omega$ and the radius of the pinning potential $R_{\rm pin}$. The parameter space is limited to frequencies $\Omega<\Omega_{\rm c}$ in order to guarantee the global stability of multiquantum states. Neither rotation of the system nor the pinning potential alone suffices to stabilize multiquantum vortices. Obviously, the minimum stabilizing radii of the pinning potential could be further lowered by increasing the rotation to frequencies $\Omega>\Omega_{\rm c}$. However, the situation would be substantially more complicated in that regime due to the possible existence of stable states lacking rotational symmetry.}
\label{Fig3}
\end{figure}

The computed stability phase diagram for multiquantum vortices in terms of the angular rotation frequency and the radius of the pinning potential is depicted in Fig.~\ref{Fig3}. Each phase covers an area of the parameter space where a multiquantum vortex with given vorticity $m$ is the minimum configuration of the energy functional. The phase boundaries are obtained by finding for each value of $R_{\rm pin}$ the rotation frequencies $\Omega$ for which multiquantum vortices with successive vorticities have equal energy, see Fig.~\ref{Fig1}. Specifically, the dashed line in Fig.~\ref{Fig2} corresponds to the vertical line at $\Omega=\Omega_{\rm c}$ in Fig.~\ref{Fig3}.

Neither the rotation of the system nor the pinning potential alone suffices to stabilize multiply quantized vortices as is readily seen from Fig.~\ref{Fig3} ---a conclusion which applies equally well outside the parameter space covered. As the width of the pinning potential is increased, more circulation quanta `fits' inside the pinning potential for a given rotation frequency. Similarly, as the rotation frequency is increased for a given pinning potential, the system can lower its energy by nucleating more circulation quanta in the multiquantum vortex located in the pinning potential. Evidently, the minimum radii of the pinning potential could be further decreased by increasing the angular rotation frequency above $\Omega_{\rm c}$--- in other words each phase displayed extends beyond $\Omega=\Omega_{\rm c}$, although the present study does not cover that regime.  

\subsection{Local stability}
The remaining question is the local energetic stability of multiquantum vortices, which is determined by the sign of the lowest quasiparticle excitation energy of the condensate. For singly quantized vortices in nonrotating systems, there exists at least one anomalous mode, \emph{i.e.}, a negative-energy eigenmode with positive norm, within the Bogoliubov prescription. By rotating the system, those quasiparticle states may be lifted to positive energies \emph{w.r.t.} the condensate energy, implying local stability of the vortex state. Such stabilization by pure rotation is not possible, however, for unpinned multiquantum vortices as discussed below.

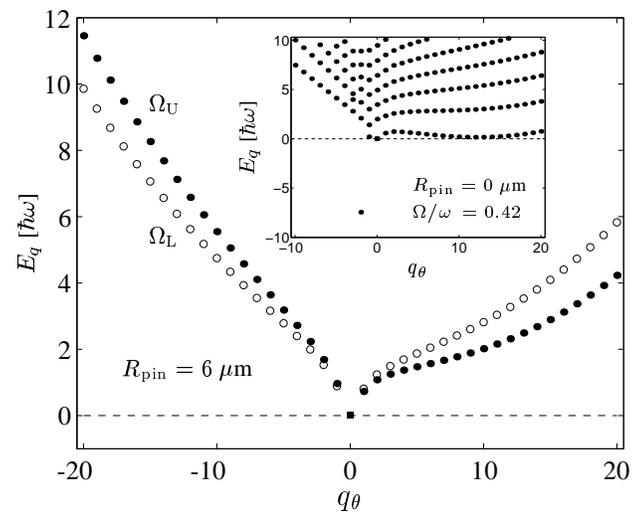
\begin{figure}[!h]
\caption{Lowest quasiparticle excitation energies for a stabilized ($R_{\rm pin}=6$ $\mu$m) doubly quantized vortex line as functions of the quasiparticle angular momentum quantum number $q_\theta$ for rotation frequencies $\Omega_{\rm L}=0.15$ ($\circ$) and $\Omega_{\rm U}=0.23$ ($\bullet$). The energies are measured relative to the condensate energy ({\tiny $\blacksquare$}). For $m>1$ the unpinned vortex cannot be made locally stable by simply rotating the system, as is demostrated by the inset for $m=2$. The anomalous mode persists below the condensate energy and further increase of rotation frequency would bring about even more anomalous excitations for higher values of the angular momentum.} 
\label{Fig4}
\end{figure}

The inset in Fig.~\ref{Fig4} displays the energies of the lowest quasiparticle states for an unpinned doubly quantized vortex line. The system is rotated at the angular frequency $\Omega/\omega=0.42$, but the anomalous mode at $q_{\theta}=-2$ lies far below the condensate energy. Further increase of rotation results in a growing number of negative-energy excitations at higher values of angular momenta. The above analysis generalizes to higher values of $m$ as well. And, the conclusion is that unpinned $m>1$ vortices are locally unstable even in the presence of a rotating drive \cite{Garcia-Ripoll1999a}. 

However, the use of a pinning potential changes the situation, and multiquantum vortices can indeed be also locally stabilized. In the main frame of Fig.~\ref{Fig4} is shown the lowest excitation energies for stable $m=2$ vortex at rotation frequencies $\Omega_{\rm L}=0.15$ ($\circ$) and $\Omega_{\rm U}=0.23$ ($\bullet$). The anomalous mode has disappeared from the spectrum as a consequence of the pinning, and the system is locally energetically stable even well beyond $\Omega_{\rm U}$. The two-fold purpose of the pinning potential is thus to lift the anomalous vortex core modes to positive energies, and to lower free energies of multiquantum vortices, rendering them locally and globally stable, respectively. 

\section{Discussion}
In conclusion, we have studied multiply quantized vortices and their energetic stability in dilute Bose-Einstein condensates using the Bogoliubov approximation. Energies of different multiquantum vortex configurations were computed and compared with each other in order to find the globally stable minimum-energy states. The analysis was restricted to rotationally symmetric states only by studying such rotation frequencies for which the nucleation of vortices outside the volume of the pinning potential is energetically hindered. We discussed both the local and global energetic stability of multiquantum vortex states and presented a phase diagram for their stabilization in terms of the radius of the pinning potential and the angular rotation frequency.

In this paper, it was shown that a combination of a pinning potential and external rotation of the system facilitates the existence of multiply quantized vortex states in harmonically trapped BECs. Such pinning of vortices, which could be accomplished, \emph{e.g.}, with an additional laser beam, has often been suggested but remains to be realized in the experiments \cite{Matthews1999b,Inouye}. Based on the above analysis, we suggest that such multiquantum vortices could be created using available experimental techniques.

\begin{acknowledgments}
We thank the Center for Scientific Computing for computer resources,
and the Academy of Finland and the Graduate School in Technical Physics
for support. 
\end{acknowledgments}

\newpage 
\bibliography{bibfile}

\end{document}